\documentclass[useAMS,usenatbib]{mn2e}
\usepackage{graphics}
\usepackage{epsf}
\usepackage{subfigure}

\begin{document}


\title[Cubic Polynomial Light Curve Fitting]{Approximating RR Lyrae light curves using cubic polynomials}
\author[Reyner et al.]{S. Reyner$^{1}$\thanks{Email: steven.reyner@oswego.edu}, S. M. Kanbur$^{2}$, C. Ngeow$^{3}$ and C. Morgan$^{1}$ 
\\
$^{1}$Department of Mathematics, SUNY Oswego, Oswego, NY 13126, USA
\\
$^{2}$Department of Physics, SUNY Oswego, Oswego, NY 13126, USA
\\
$^{3}$Graduate Institute of Astronomy, National Central University, Jhongli City
, 32001, Taiwan (R.O.C.)}

\date{Received XX XXX 2010 / Accepted XX XXX 2010}

\maketitle

\begin{abstract}
In this paper, we use cubic polynomials to approximate RR Lyrae light curves and 
apply the method to $HST$ data of RR Lyrae stars in the halo of M31. We compare our results to the
standard method of Fourier decomposition and find that the method of cubic polynomials eliminates virtually all
ringing effects and does so with significantly fewer parameters than the Fourier technique.
Further, for RRc stars the
parameters in the fit are all physical in the sense that they can, in principle, be related to pulsation physics.
Our study also reveals a number of additional periodicities in this data not found previously: we find 23 RR$c$ stars, 29 RR$ab$ stars and
3 multiperiodic stars.

\end{abstract}



\section{Introduction}

Even though a number of microlensing projects have yielded RR Lyrae light curves with excellent
phase coverage \citep[see, for example,][]{soz08}, phase coverage is more of a problem when studying RR Lyraes in external galaxies.
In order to overcome this, a number of methods have been developed to use the observed data to get a smooth approximation
to the actual RR Lyrae light curve that captures any real physical bumps or dips and eliminates any that are caused by
numerics.

The traditional method applied has been Fourier decomposition wherein the observed data are fit by an expression of the form
$$m(t) = A_0 + \sum_{k=1}^{N}A_k\sin (k{\omega}t + {\phi}_k). \eqno(1)$$
Here $N$ is the order of the fit, $A_0$ is the mean magnitude, ${\omega} = 2{\pi}/P$, $P$ is the pulsating period and $t$ is the time
of observation. Usually a least squares fitting procedure yields $A_0$, $A_k$, and ${\phi}_k.$ A major problem with this technique is
ringing: the Fourier curve given by equation (1) exhibits a series of unphysical bumps and dips. This can occur even when
the original data are well distributed in phase but exhibit a large scatter. This is exacerbated when there are noticable gaps in
phase coverage. The solution to this is to reduce the order of the fit but then this looses real features of the light curve.  

Another method is that of Principal Component Analysis \citep[PCA, see ][for examples]{kn04,deb09}. Here instead of
sine functions being the basis, the data itself
determines these basis functions: the resulting light curve is expressed as a sum of these basis functions. PCA has the advantage that
a very good approximation of the light curve can be realised with significantly fewer parameters than required by the Fourier method.
However, while a stringent comparison of the two methods is beyond the scope of this paper, we can say that, at least in the case of RR$c$ stars,
the parameters of a fit using the methods developed here do have some physical interpretation.

\citet{ak94} studied the method of cubic splines in approximating variable star light curves. A cubic spline is a series of cubic
polynomials pieced together such that the intersection of two such polynomials is required to be continuous up to the second
derivative. Our method uses cubic polynomials but does not require continuity up to the second
derivative. Looking at figure 4 in \citet{ak94}, the graphs for LCB12, LCR12, we again see
ringing as a result of having 16 parameters: as our results show, this is a clear reason why cubics polynomials as opposed to
cubic splines are to be preferred in approximating RR Lyrae light curves.

One motivation for this study is that requiring continuity up to the second derivative seems too stringent. Pulsation shocks are dramatic events with sudden
reversals. There is no reason to suppose any fitted curve is continuously differentiable to the second degree.
This is particularly true of fundamental mode RR Lyraes at phases close to minimum light where the star suddenly starts to get brighter.
In this paper we examine the use of cubic polynomials to fit the light curves of RR Lyraes by requiring continuity only up
to the first derivative.

\section{Preliminaries}

We define $t \bmod P$ (for any real number $t$ and positive real $P$) to be that positive number $x$ satisfying both
$0\le x < P$ and $(t-x)/P$ is an integer. We are interested in approximating data points: $(t_1,y_1),\ldots, (t_n,y_n)$ by
a periodic function $y=f(t)$ of period $P$. The residuals from our fit are $r_i = y_i - f(t_i).$ We define $PD$ to be the proportion of the period when the
luminosity is decreasing (which, when magnitudes are used, results in the proportion of the period when the $y_i$ are increasing).
Let $\bar{y}$ be the average of the $y_i.$ Define the Total Sum of Squares, $SST$, to be
$$SST = \sum_{i=1}^n[\bar{y} - f(t_i)]^2,$$
and the Error Sum of Squares, $SSE$, as
$$SSE = \sum_{i=1}^n r_i^2.$$
The quantity $R^2$, is defined as,
$$R^2 = {{(SST-SSE)}\over{SST}},$$
and is the portion of the variation of the $y_i's$ explained by our model $f(t)$. The adjusted $R^2(adj)$, denoted by $RA$, is
$$R^2(adj) = 1 - (n-1)(1-R^2)/(n-r-1),$$
where there are $r$ parameters.
Then the best fit is from that combination of parameters such thet $SSE$ is a minimum. 

For any function which is twice differentiable except for a finite set of points, we define the total bending ($TB$) as follows:
divide the domain ($[0,P]$ if periodic with period $P$) into intervals so that on each interval, the function is only
concave up or down. On each interval, the bending is the angle between the tangent lines at the two endpoints (not through the vertical) and at 
each nondifferentiable point, the angle between the left and right sided tangent lines. $TB$ is the sum of all these angles.
This provides a good measure
of "ringing".

We use an $F$ test \citep{w80} to test for the significance of having a more/less complex model. This $F$ statistic is
$$\frac{[SSE(N) - SSE(A)]/[df(N)-df(A)]}{SSE(A)/df(A)},$$ 
where $SSE$ is the error sum of squares, $N$ and $A$ stand for null hypothesis (less complex model) and alternative hypothesis (more complex model) respectively. The expression $df$ stands for degrees of freedom which is the number of data points minus the number of parameters.

\section{Our Approximation}

Our goal is to obtain good approximations which reflect the true shape of the light curve, yet are simple and without ringing or other
anomalies. Our method is based on the observation than an increasing or decreasing portion of $\sin(t)$ function can
be remarkably well approximated by a cubic polynomial. We note also that a cubic polynomial is uniquely determined by the $y$ coordinate and the
slope at two points.

RR Lyrae light curves come in basically two varieties: types $ab$ and $c$ corresponding to fundamental and overtone modes respectively. For the
overtone $c$ type, both increasing and decreasing portions are roughly half the period of a sine curve (of different periods) and each can be 
approximated by a cubic. For the fundamental $ab$ type, the increasing portion is roughly half of a period of a sine curve, while the
decreasing part is similar to the bottom half of the decreasing half of a sine curve, though near the minimum, there is a noticeable dip
before it starts to increase again.

In this paragraph, we outline the method in general terms and describe the details in the following paragraphs. Essentially
approximate RR Lyrae light curves by either 2 cubics or 3 cubics. When we use 2 cubics, for example when trying to model RRc curves, the
parameters of our fit are the Period, Shift (phase point at which first observation occurs), the Maximum and Minimum and the Proportion of the
Curve that is Decreasing and the slope at maximum, that is a total of 6 parameters. In this case, it is clear that these
parameters have a physical meaning. We choose these parameters in order to minimize the $SSE.$
In this case the fitted curve is continous as is its derivative except perhaps at maximum. When we use 3 cubics, we choose 4 points with the understanding
that the first and last points are the same and 3 time intervals. Note the sum of these three time intervals is the period. We have the
$x$ and $y$ values at these points and the slopes. This is a total of 9 parameters together with the shift as defined in the case of using 2 cubics. We choose
these 10 parameters to minimize the $SSE$ as before.

With this in mind, we define two different piecewise defined functions and their periodic extensions. $S_1(t)$ is defined on
$0 \le t \le T_1$ to be that cubic which passes through the points $(0,M)$, and $(T_1,m)$, with derivative equal to
$D$ at $t=0$ and with zero derivative at $T_1.$ Here $M$ and $m$ are the maximum and minimum, respectively of the curve to be fitted.
Furthermore, $S_1(t)$ is defined on $T_1 \le t \le P$ to be that cubic which 
passes through the points $(T_1,m)$ and $(P,M),$ with derivative equal to zero at $t=T_1$ and $P.$ The periodic
extension $S_1(t \bmod P)$ is said to be of type $2C.$ This is continuous since $S_1(0) = S_1(P).$ Note that for
sinusoidal curves, $D$ will equal 0 ($2C$ will be differentiable) while for fundamental $ab$ type curves, $D$ will be positive and
$2C$ will not be differentiable. In all that follows, when we approximate a type $c$ by $2C$, we require $D=0$ and denote this by $2C-0.$
Our parameters here are period, shift (how far into the period the first data point is), $D$, $m$, $M$, and $PD.$ Again each of these
parameters has some physical meaning.
Similarly $S_2(t)$ is piecewise defined by dividing $[0,P]$ into three intervals, using a cubic on each and requiring continuity
and differentiability where any two cubics meet. As before its periodic extension is of type $3C.$ Our parameters here are
period, shift, the two time values where continuity is required between two cubics, three $y$ values and three slopes.
Because of the simplicity of both $2C-0$ and $3C$, there is no ringing. Consequently, no noise is added when obtaining residuals.
Furthermore, having fewer parameters improves the power of an $F$ test.

In practice, we compute the mean and standard deviation for the original data. Points beyond 2.5 standard deviations away from the
mean are considered outliers and omitted. We need good initial guesses to best approximate out data. For $2C$, our initial approximation
comes from the Fourier series of 7 terms and $D=0.$ We use $T_1=PD*P$ and replace each data point by $(t_i,y_i)$ by
$(t_i-shift,y_i).$ We are then in a position to minimize $SSE.$ From the Fourier fit, we find the maximum as the point (shift, $M$),
followed by the minimum as (shift + $T_1$, m). For $3C$, our initial approximation uses $2C$ and we define
$T_0$ to be between $60\%$ and $90\%$ of $T_1$ (see later) and use the intervals $0\le t\le T_0$,
$T_0 \le t \le T_1,$ and $T_1 \le t \le P$ for our three cubics. $P$ is obtained by maximizing the power function. We obtain both the
$y$ coordinate and slope at $0$, $T_0$, $T_1$ from $2C$ from which we obtain $3C.$ For $3C$, we define $T_0$ as $60\%$, $70\%$, $80\%$ and then $90\%$ of $T_1$.
For each case, we find the minimum $SSE$ and keep the best (smallest $SSE$) set of parameters. We continue our removal of outliers.
For the residual, removing outliers based on the Fourier series is not appropriate because of ringing. We look at residuals using both $2C$ and $3C$.
For each, we compute the mean and standard deviation. We remove any points more than 3 standard deviations out using both criteria. When
comparing various approximations, we compare all on this same final data set.

We minimize $SSE$ by looping through the parameters with successively smaller step sizes, $s$. For a fixed parameter, in
addition to having the current value of $SSE,$ we evaluate $SSE$ at this parameter plus $s$, and at this parameter minus $s$. Finally
we fit these three points by a quadratic polynomial, find its minimum and evaluate $SSE$ at this point. Of the current four estimates of
our parameter, we select the one giving the smallest $SSE$ as the parameter's new value. We continue this until we have a good approximation of
a relative minimum.

\section{Test Data}

\begin{figure}
 \vspace{0cm}
\centering
   \epsfxsize=8.0cm{\epsfbox{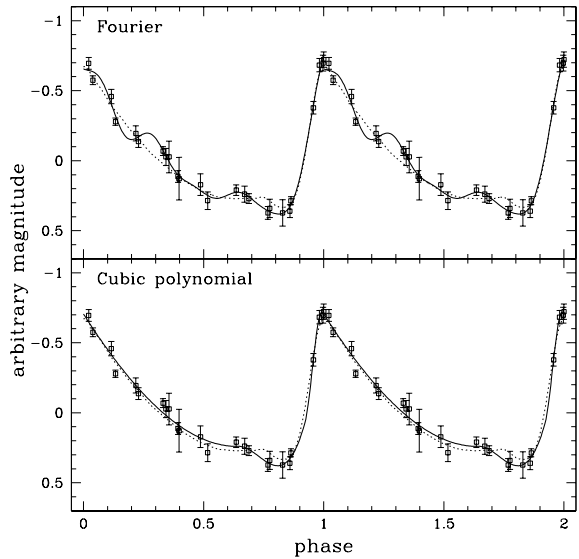}}
    \caption{Results of fitting to data (open squares) drawn from a known (dashed curves) function using Fourier (top panel, solid curve)
and cubic polynomial (bottom panel, solid curve) methods.}
\label{UNIQUE_LABEL}
\end{figure}

We first test our method on a known function, where the known function is taken from the RR Lyrae light curve templates developed by \citet{lay98}.
We draw a random number of points from this function and independently add Gaussian noise to each phase point.
Specifically, we add noise normally distributed with mean zero and standard deviation 0.1 to the synthetic data. In our tests, this
synthetic data is not a "train" of data but just covers approxiametly one period.
We then try to reproduce the original curve using our cubic polynomial method and the traditional Fourier technique.
Figure 1 presents our results. Here the known function (chosen to resemble a typical RR$ab$ type light curve) is the
dashed curve.
The open squares with error bars
are the points drawn randomly from this function and to which Gaussian noise has been added independently. The solid
dark lines represent the Fourier and cubic polynomial fits (top and bottom panels respectively) to these open square points.
The curves are plotted as a function of phase, going from 0 to 1. However, we allow the two methods to "rediscover" this periodicity.
The Fourier fit of order 6 yielded considerable ringing, a period of 1.14
and had a $SSE$ of $0.027$. We emphasize that by a period of 1.14, we do not mean a period of 1.14 days, but that this signifies a change in the period as
reported by the Fourier method when compared to the period of the original template curve from which the data were drawn. A reported period of 1 signifies
no difference between the estimated and original period.
Approximating by a pair of cubics ($2C$), we obtained a slightly different period of 0.997, $D=2.93$,
$PD=0.905$ and $SSE=0.050$. These values of $D$ and $PD$ both indicate Bailey type $ab$. Using a $3C$ (differentialble) approximation, we obtained
a period of 1.00004 and $SSE=0.04$. We see that our method does a very good job at mimicking the known function and
has little to no ringing. In contrast, a Fourier fit to the same points produces noticeable and significant ringing. Further, the period
obtained by the cubic polynomial method matches exactly that of the original curve.  

Next we tested our method on real data.
The data were taken from \citet{br04} and consist of $HST$ observations of RR Lyrae stars in the Andromeda Halo - along the
southeast minor axis of M31, about $51'$ from the nucleus. The data are available at two wavelengths, F606W and F814W.
In what follows we report results based on both bands (F606W and F814W are referred to as the
first and second bands respectively). \citet{br04} used a fast algorithm based on the
Lomb-Scargale periodogram \citep{sc82} to search for periodicities in their time series data after data reduction and photometry.
\citet{br04} analyzed these data and found
169 variables of which 55 were clearly RR Lyraes. Of
these 55 stars, \citet{br04} classified 29 as RR$ab$, 25 as RR$c$ and 1 as RR$d$.  It is the data for these 55
stars which we discuss in this paper. Note we start with the photometry for these 55 stars as published in \citet{br04}.
As pointed out by the referee, intrinsic precision in this dataset is about 0.03 and 0.04 in $V,I$ respectively. This measurement error
is not accounted for in our fits but for the purposes of this paper it is appropriate since
we are presenting a differential comparison between our method and
that of Fourier series.
Our data are not corrected for reddening. \citet{br04} found a ratio of RR$c$ to RR$abc$ of 0.46,
mean periods of RR$c$ and RR$ab$ stars of 0.316 and 0.594 days respectively. In figures 2-4, the label "normalized magnitude" just
refers to the fact that the magnitudes are scaled to lie between 0 and 1.

We note that these are HST data and as such somewhat immune to the 1-day aliasing problems arising when RR Lyraes are observed from the ground.
The sampling rate was 250 exposures over a 41 day period \citep{br04} with a cadence that should be random enough to offset other aliasing problems.
Observations in the two bands were made at slightly different times: this again helps to counteract aliasing.

\section{Results}

\begin{figure}
 \centering 
 \epsfxsize=8.0cm{\epsfbox{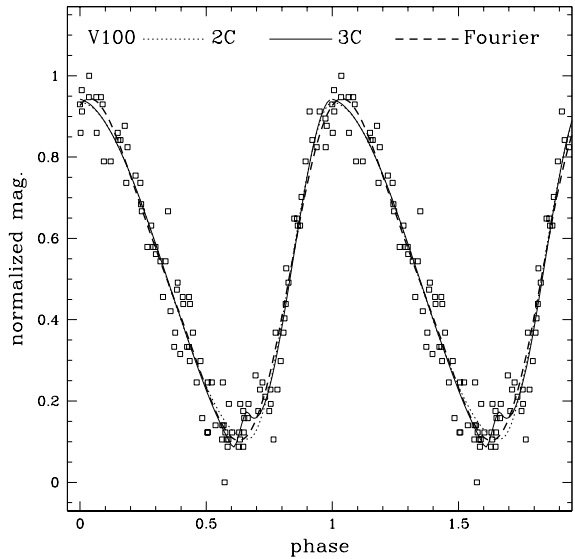}}
 \caption{Results for RR$c$ star V100: open squares are data, thin/thick dashed curves uses $2C$ and Fourier respectively, solid curve uses $3C.$ The $y$ axis is scaled such that the range of magnitudes goes from 0 to 1.}
 \label{UNIQUE_LABEL}
\end{figure}

A major finding of our work is that when using the cubic polynomials method on the dataset mentioned,
we find 23 RR$c$ stars with a mean period of 0.312, 29 RR$ab$
stars with a mean period of 0.594.
This leads to a ratio of RR$c$ to RR$abc$ stars very similar to previous work: 0.442.
This ratio is lower than that reported in \citet{br04} because there were two $RRc$ stars which were reclassified as $RRd$ in
our work.
In almost all cases, the periods discovered by our method is very close to that published by \citet{br04}.
An intriguing result is that our method reveals significantly more multimode stars than previously discovered
and we discuss this later in this section.

First overtone stars all have periods less than 0.39 whilst fundamental mode stars have periods greater than 0.44.
A nice result is that the type $c$ stars all have $0.51 < PD < 0.76$ while all type $ab$ stars have $PD > 0.76.$ We observe three other
imperfect tests for Bailey type. Type $c$ has $D$ (first band) less than 0.76 except for V76 and V58. $PD$ (second band) is 
less than 0.76 except for V120; $D$ (second band) is less than 0.25 except for V40 and V120. Also,
type $ab$ has $D$ (first band) greater than 0.25 except for V78; $PD$ (second band greater than 0.76
except for V122; and $D$ (second band) is greater than 0.25 except for V66, V71 and V82.
In every case, both bands have $PD$ greater than 0.51 with two exceptions, both of which are in the second band where there is
a generally more noise. For V76, $PD=0.36$ while for V157, $PD=0.48.$ Generally, the $PD$'s of the two bands correlate nicely, as do the $D$'s.
Further, type $c$ all have
$D < 0.2$ while all type $ab$ except V78 have $D > 0.5.$ This provides a good way to distinguish between types $ab$ and $c.$
Type $c$ can be approximated about equally well by a Fourier series of order 2-4, or $2C$ or $3C$. The advantages of using $2C$
are its simplicity, minimal $TB$ and using only parameters that have physical meaning. Further we can generate $PD$ and $D$ whose
importance has already been established.

The 23 type RR$c$ data sets can be summarized on average as follows. The average RA for
$2C$ and $3C$, we call $R2c$ and $R3c$, respectively. A similar quantity for an order 2 Fourier series is labelled as $Rf2c.$ Likewise, $TB$
for $2C$ is named as $TB2c$ and so on. We have $R2c=0.906,R3c=0.907$ and $Rf2c=0.902$ while $TB2c=5.2,$ $TB3c=8.2$ and $TBf2c=5.2$.
While $3C$ may give slightly better approximations the extra bending strongly suggests it is not worth the effort. Fourier series give
somewhat worse approximations than $2C$ with no reduction in bending. Finally, $2C$ is simplest and only involves parameters
with physical meaning which clearly makes it superior. Figure 2 displays a typical example of type $c$ using star V100. For this
star, $R2=0.951$, $R3=0.951$, $Rf2=0.950$, $TB2=5.5$, $TB3=10.5$ and $TBf2=5.5$.

The 29 type $ab$ stars can be summarized on average as follows.
Using similar nomenclature to that specified for the RR$c$ stars above, we have
$R2ab = 0.954$, $R3ab = 0.962$ and $Rf8ab = 0.962$ while $TB2ab = 5.4$, $TB3ab=8.4$ and $TBf8 = 16.4.$ Since $3C$ gives about as good an approximation as Fourier
series with 8 terms but with much less bending and fewer parameters, we prefer $3C$. Comparing $R2ab$ to $R3ab$ we see that
$RA$ has gone from 0.954 to 0.962, which means the ratio of errors, $(1-R2ab)/(1-R3ab) = 1.21$: a $21\%$ reduction in error so $3C$ is
clearly superior. The increase in bending simply means we are twisting more to fit the data - as can be seen especially well in figures 1 to 3.
Figures 1 to 3 are typical examples of the sort of approximations possible with cubic polynomials.
Further,
figure 3 presents typical examples of type $ab$ using V57 and V136. V57 (left panel of figure 3) is an example (of 6 or 7 stars) where the
decreasing portion seems to momentarily increase before a final dip to the minimum, while V136 (right panel of figure 3) does not show this behavior.
For V57 we have $R2=0.953$, $R3=0.969$, $Rf8=0.970$, $TB2=5.8$, $TB3=10.4$ and $TBf8=16.2$, and 
for V136 we have $R2=0.933$, $R3=0.937$, $Rf8=0.938$, $TB2=4.7$, $TB3=7.1$ and $TBf8=13.8$.

The proportion decreasing using $3C$ is not the same as $PD$ (using $2C$). However, it is usually within $0.001.$
The period obtained by $2C$ and $3C$ are usually within $0.0001$, while they differ from the period using
Fourier series by perhaps $0.001$. Worse yet, the optimal period for a Fourier series depends on the order of the series.

There are three stars, listed in Table 1, which have two periods. These were analyzed as follows. We removed outliers in the data as before.
We then fit by $2C$ or $2C-0$ depending on type, and removed outliers based on their residuals.
This gave an initial period equal to $prd1$, $RA=RA1$ and $SSE=SSE1.$
We then obtained the next period and its power $=pwr.$ We subsequently refit by $2C-0$ if necessary, and then we fit the residuals by
another $2C-0$ and finally obtained
a combined best fit (with both $D=0$). This gives $prd2$, $RA2$ and $SSE2$, from which we calculated an $F$ statistic. These are all listed in Table 1.
For each star, the first and second lines correspond to the first (F606W) and second (F814W) bands, respectively.
The $F$ statistic tells us that we are more than
$99.95\%$ certain that the second $2C-0$ is significant and so the second period is significant. A different approach is based on Scargle's analysis.
Using his equation (18), if it is possible to select $N$ possible periods a priori, then in order to conclude that, with greater than $99\%$ certainty,
the best one
is valid, the power for more than one must exceed the threshold $-\ln [1-0.99^{(1/N)}].$ However there is no way of knowing in advance what the true period is.
If we assume it lies between 0.250 and 0.800 and round to three decimal places, there are 551 possibilities which gives a threshold of 10.9. Each of the
powers listed in Table 1 is above this except for V95 in the second band. If a second period is present it should be present in each band.
We conducted an extensive search in the first band which has higher amplitude and then checked some periods near the predicted period - the second power
was conclusive. For a number of stars, using only one band, we discovered we could obtain a much better approximation using two periods differing by only about
0.001. We assume these are anomolies and ignore them though more data might lead to different results. Several stars had significant evidence
of a second period in one band but not in the other and these were ignored.
These stringent criteria leave the three stars in Table 1 with two periods. In
each case the period ratio is 0.75. Figure 4 displays a graph of V1 over 4 primary periods which equals 3 secondary periods. This also presents the
interaction of the two periods. These three stars have primary periods between 0.353 and 0.383.

\begin{figure*}
  \vspace{0cm}
  \hbox{\hspace{0.5cm}
    \epsfxsize=8.0cm \epsfbox{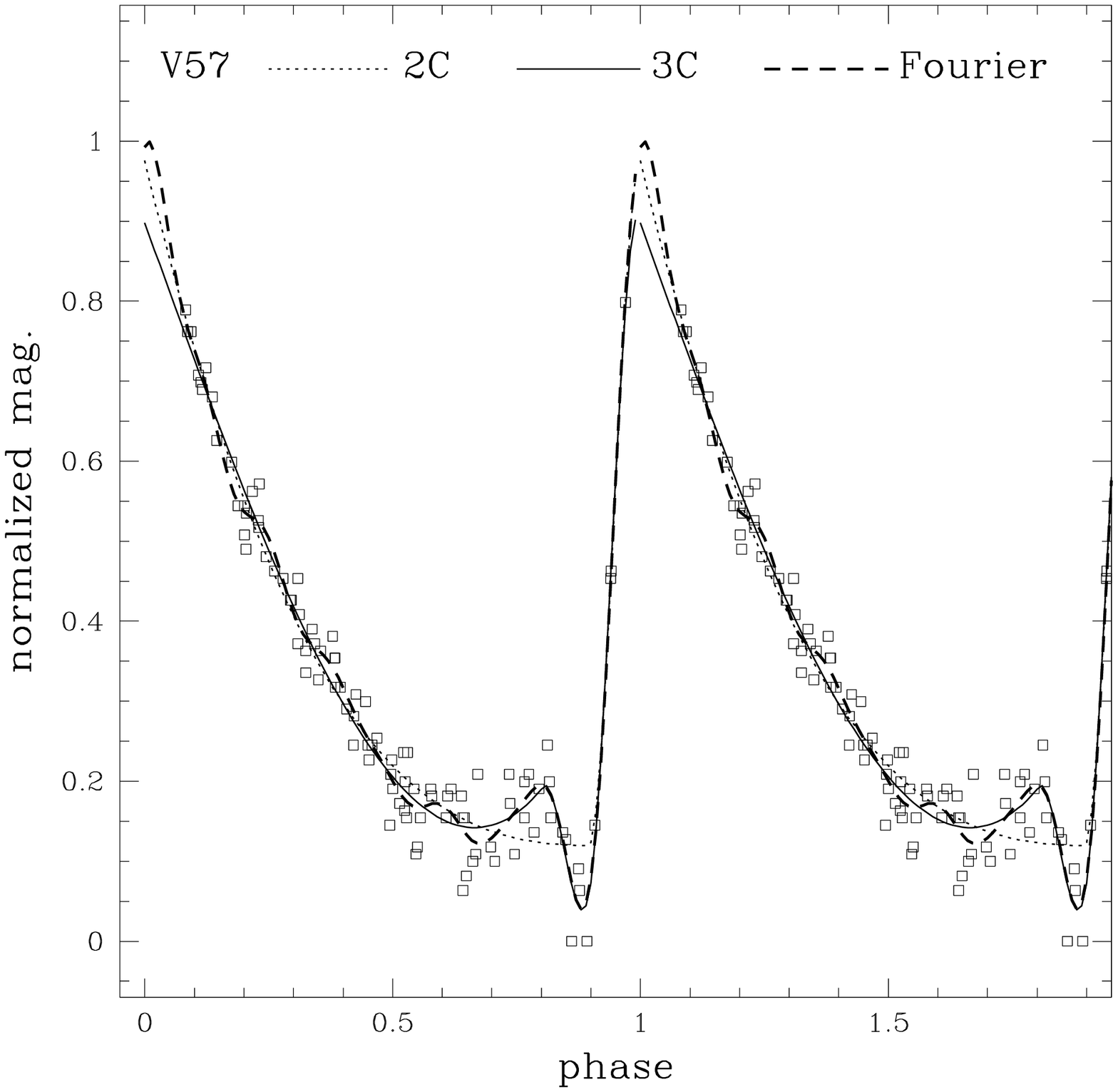}
    \epsfxsize=8.0cm \epsfbox{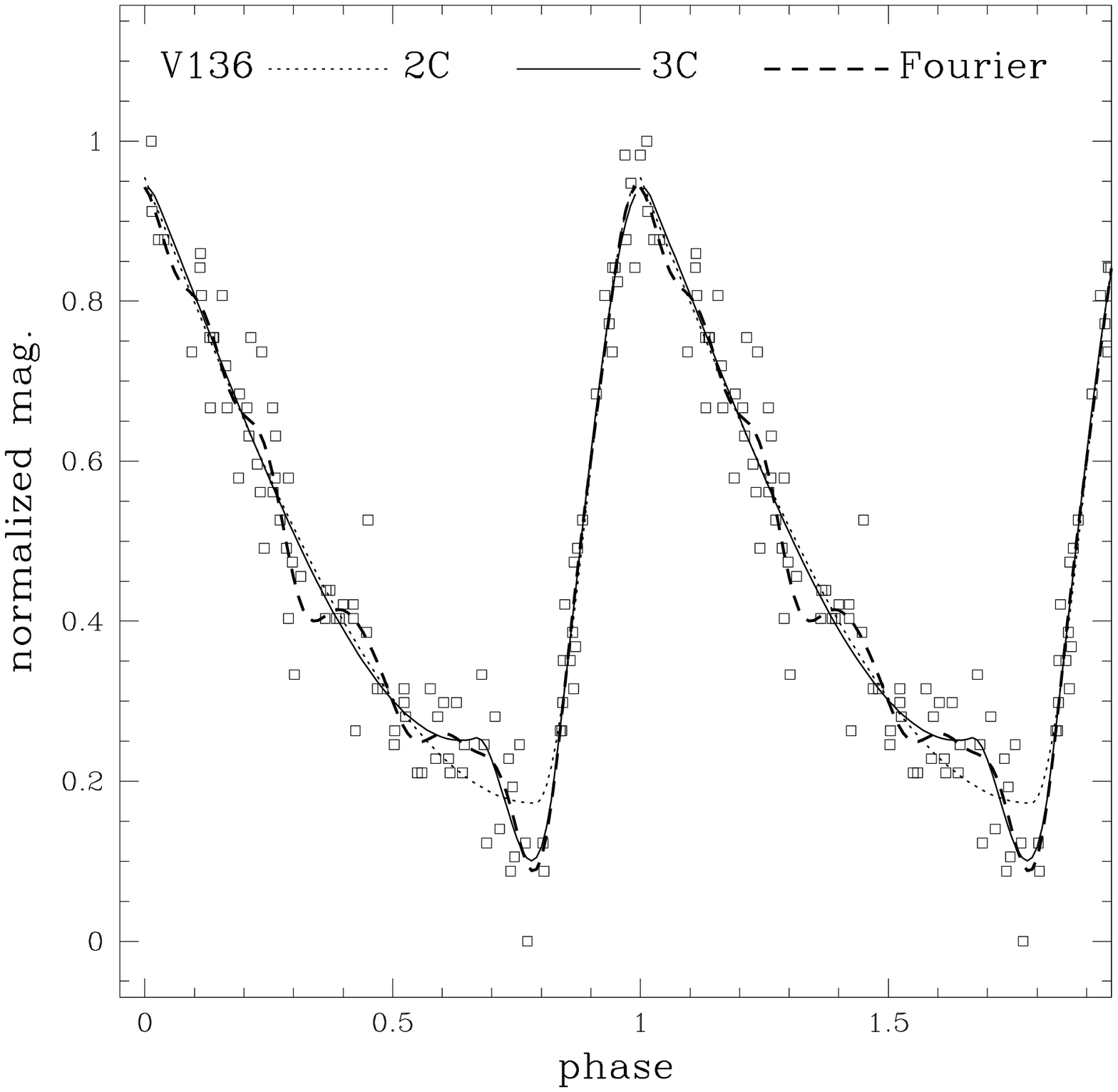}
  }
  \vspace{0cm}
  \caption{Results for RR$ab$ stars V57 (left panel) and V136 (right panel): open squares are data, thin/thick dashed curves
uses $2C$ and Fourier, solid curves uses $3C$.}
  \label{UNIQUE_LABEL}
\end{figure*}

\section{Maximum and Minimum Light Colors}

Recent work has focused on the properties of RR Lyraes at minimum light has a possible way to estimate reddening \citep{lub77,cle95,kf05}.
The theoretical basis for this has been established by \citet{skm93}, \citet{kan95} and \citet{kp96}. These authors
showed the importance of Period-Color (PC) relations at maximum light. Cepheids have flatter PC relations at maximum light and definite
relation at minimum light such that higher amplitude Cepheids are driven to cooler and hence redder colors. In the case of RR Lyraes this
is reversed with a flat PC relation at minimum light and a discernable relation at maximum light. Figure 5 presents PC relations
at maximum and minimum light for the \citet{br04} data calculated using both Fourier series (open circles) and cubic polynomials (solid
black squares) to approximate the data. Firstly, we see broad support for the contention that PC relations at minimum light are much
flatter than those at maximum light. Secondly, as pointed out by the referee, we notice somewhat tighter and flatter relations are
present for the PC relation at minimum light when using a cubic polynomial fit.

\section{Conclusions}

We have found a new way to approximate the light curve of an RR Lyrae star by fitting cubic polynomials to the data. This method can
fit the data with fewer parameters than Fourier series and suffers virtually no ringing. It can also estimate periodicities in the data.
When we apply this method to RR Lyrae data in the Andromeda halo, we find, in addition to the multiperiodic star V90 reported by \citet{br04},
an additional 2 other multiperiodic stars (V1 and V95, previously classified as type $RRc$) in the data sample: here we require this multiperiodicity to be present in both
bands.
Then the ratio of the number of RR$c$ stars to the ratio of the number of RR$abc$ stars is 0.442 - as opposed to \citet{br04} who found a ratio of 0.462.
In this ratio, \citet{br04} do not count RRd stars in either numerator or denominator. The ratio of number of $RRc$ to
number of $RRab$ where we include $RRd$ stars together with the $RRc$ stars is in this case is 0.473.

\begin{table*}
\centering
\caption{Stars with multi-periodic components}
\label{NOLABEL}
\begin{tabular}{ccccccccccc}\hline
V  & Band & Period1 & $RA1$ & $SSE1$ & Period2 & $RA2$ & $SSE2$ & $pwr$ & $F$ & Number of points\\
\hline

1  & F606W & 0.3815 & 0.856 & 0.252 & 0.5104 & 0.940 & 0.096 & 13.6 & 26.3 & 74  \\
1  & F814W & 0.3816 & 0.779 & 0.227 & 0.5108 & 0.865 & 0.133 & 16.4 & 14.6 & 91 \\
90 & F606W & 0.3533 & 0.823 & 0.579 & 0.4742 & 0.919 & 0.252 & 18.8 & 27.3 & 93\\
90 & F814W & 0.3533 & 0.572 & 0.616 & 0.4747 & 0.809 & 0.263 & 26.2 & 30.2 & 99\\
95 & F606W & 0.3616 & 0.781 & 0.429 & 0.4855 & 0.910 & 0.164 & 19.0 & 36.4 & 99\\
95 & F814W & 0.3614 & 0.697 & 0.379 & 0.4855 & 0.744 & 0.309 &  9.7 & 6.0 & 115\\
\hline
\end{tabular}
\end{table*}

\begin{figure}
 \centering 
 \epsfxsize=8.0cm{\epsfbox{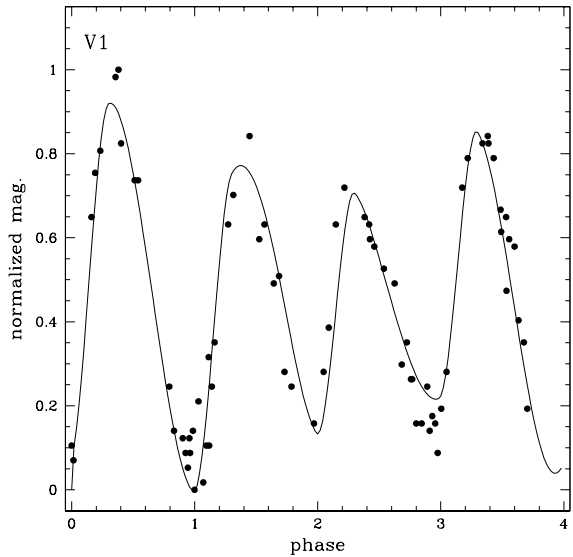}}
 \caption{Results for multiperiodic star V1: an example of a multiperiodic star: the two dominant periods are $P_1 = 0.382$ and
$P_0 = 0.510$ with a period ratio of 0.75.}
 \label{UNIQUE_LABEL}
\end{figure}

\begin{figure}
 \centering
 \epsfxsize=8.0cm{\epsfbox{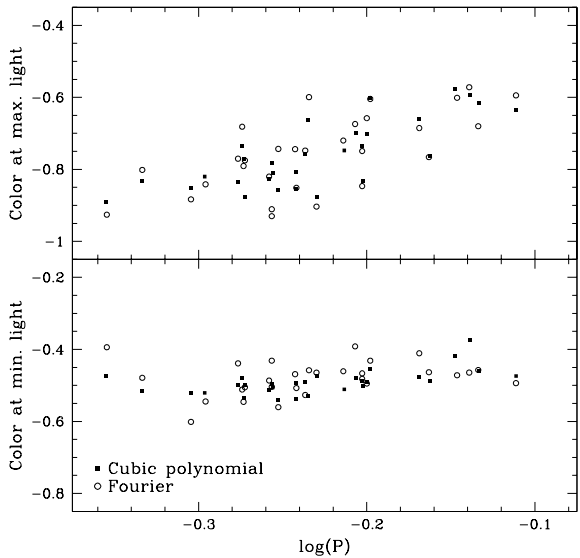}}
 \caption{PC results at maximum (top) and minimum (bottom) light using a sixth order Fourier fit (open circles) and cubic
polynomials (solid black squares).}
 \label{UNIQUE_LABEL}
\end{figure}

\section*{acknowledgments}
CCN thank the funding from National Science Council (of Taiwan) under the contract NSC 98-2112-M-008-013-MY3.
The authors thank Karen Kinemuchi for helpful discussions during the preparation of this manuscript.
The authors also thank the referee, Jan Lub, for very useful comments.




\begin{thebibliography}{}

\bibitem[\protect\citeauthoryear{Akerlof et al.}{1994}]{ak94} Akerlof, C. et al., 1994, ApJ, 436, 787

\bibitem[\protect\citeauthoryear{Brown et al.}{2004}]{br04} Brown, T. M., et al., 2004, AJ, 127, 2738

\bibitem[\protect\citeauthoryear{Clementini et al.}{1995}]{cle95} Clementini, G., et al, 1995, AJ, 110, 2319

\bibitem[\protect\citeauthoryear{Deb \& Singh}{2009}]{deb09} Deb, S., \& Singh, H. P., 2009, A\&A, 507, 1729 

\bibitem[\protect\citeauthoryear{Kanbur}{1995}]{kan95} Kanbur, S. M., 1995, A\&A, 297L, 91

\bibitem[\protect\citeauthoryear{Kanbur \& Phillips}{1996}]{kp96} Kanbur, S. M. \& Phillips, P., 1996, A\&A, 314, 514

\bibitem[\protect\citeauthoryear{Kanbur \& Mariani}{2004}]{kn04} Kanbur, S. M. \& Mariani, H., 2004, MNRAS, 355, 1361

\bibitem[\protect\citeauthoryear{Kanbur \& Fernando}{2005}]{kf05} Kanbur, S. M. \& Fernando, I., 2005, MNRAS, 359L, 15
 
\bibitem[\protect\citeauthoryear{Layden}{1998}]{lay98} Layden, A. C., 1998, AJ, 115, 193 

\bibitem[\protect\citeauthoryear{Lub}{1977}]{lub77} Lub, J., 1977, A\&A Suppl, 29, 345

\bibitem[\protect\citeauthoryear{Preston}{1959}]{pre59} Preston, G. W., 1959, ApJ, 130, 507

\bibitem[\protect\citeauthoryear{Scargle}{1982}]{sc82} Scargle, J. D., 1982, ApJ, 263, 835

\bibitem[\protect\citeauthoryear{Soszynski et al.}{2009}]{soz08} Soszynski, I., et al., 2009, Acta Astron., 59, 1

\bibitem[\protect\citeauthoryear{Simon et al.}{1993}]{skm93} Simon, N. R., Kanbur, S. M. \& Mihalas, D., 1993, ApJ, 414, 310

\bibitem[\protect\citeauthoryear{Weisberg, S.}{1980}]{w80} Weisberg, S., 1980, {\it Applied Regression Analysis}, John Wiley \& Sons, $1^{st}$ edition

\end{thebibliography}
\end{document}